\documentclass[aps,showpacs,showkeys,12pt,preprintnumbers]{revtex4}
\usepackage{graphicx}
\usepackage{color}
\usepackage{amsmath}
\usepackage{amssymb}
\usepackage{multirow}
\usepackage{titlesec}
\makeatletter
\renewcommand\paragraph{\@startsection{paragraph}{4}{\z@}%
            {-2.5ex\@plus -1ex \@minus -.25ex}%
            {1.25ex \@plus .25ex}%
            {\normalfont\normalsize\bfseries}}
\makeatother
\def \be  {\begin{equation}}
\def \ee  {\end{equation}}
\def \ee  {\end{equation}}
\def \bea {\begin{eqnarray}}
\def \eea {\end{eqnarray}}

\begin{document}

\preprint{ECTP-2020-04}
\preprint{WLCAPP-2020-04}
\vspace*{5mm}

\title{An almost-entirely precise empirical estimation for various chemical potentials}

\author{Hayam Yassin}
\email{hiam_hussien@women.asu.edu.eg}
\affiliation{Physics Department, Faculty of Women for Arts, Science and Education, Ain Shams University, 11577 Cairo, Egypt}

\author{Eman R. Abo Elyazeed}
\email{eman.reda@women.asu.edu.eg}
\affiliation{Physics Department, Faculty of Women for Arts, Science and Education, Ain Shams University, 11577 Cairo, Egypt}

\author{Abdel Nasser Tawfik}
\email{atawfik@nu.edu.eg}
\affiliation{Nile University, Egyptian Center for Theoretical Physics (ECTP), Juhayna Square of 26th-July-Corridor, 12588 Giza, Egypt}

\begin{abstract}

The transverse momentum spectra of the well-identified produced particles, $\pi^+$, $\pi^-$, $K^+$, $K^-$, $p$, $\bar{p}$, $K_s^0$, $\Lambda$, $\bar{\Lambda}$, $\Xi^-$, and $\Xi^+$ are analyzed in a statistical approach. From the partition function of grand-canonical ensemble, we propose a generic expression for the dependence of the generic chemical potential $\mu$ on the rapidity $y$. Then, by fitting this expression to the experimental results on the most central $p_{\perp}$ and $d^2 N/2 \pi p_{\bot} dp_{\bot} dy$, at  energies ranging from $7.7$ to $200~$GeV, we have introduced a generic expression for the rapidity dependence of $\mu$, at different energies, namely $\mu=a+b y^2$, where $a$ and $b$ is constants. We find that the resulting energy dependence $\sqrt{s_{\mathtt{NN}}}=c[(\mu-a)/b]^{d/2}$ agrees well with the statistical thermal models. We also present precise estimations of various types of chemical potentials, $\mu_B$, $\mu_S$, and $\mu_Q$, the baryon, the strangeness, and the charge chemical potential, respectively.

\end{abstract}


\maketitle
\date{\today}

%\tableofcontents
%\makeatletter
%\let\toc@pre\relax
%\let\toc@post\relax
%\makeatother

\section{Introduction}
\label{sec:into}

The high-energy experiments at the Super Protonsynchrotron (SPS) at CERN, the Relativistic Heavy Ion Collider (RHIC) at BNL and the Large Hadron Collider (LHC) at CERN have collected various evidences for the creation of the partonic matter, the quark-gluon plasma (QGP) \cite{Tawfik:2000mw,Heinz:2000ba,Gyulassy:2004zy,Heinz:2011kt,Adamczyk:2013dal,Ryu:2017qzn,Bzdak:2019pkr}. In 1999, first evuidences have been provided by SPS at CERN \cite{Tawfik:2000mw,Heinz:2000ba}. The RHIC discovery announced in 2004 confirmed this and also estimated the viscous properties of QGP \cite{Gyulassy:2004zy}. LHC \cite{Heinz:2011kt,Ryu:2017qzn} in turn confirmed SPS and RHIC discoveries which have been strengthened by STAR-BES, at RHIC energies \cite{Adamczyk:2013dal,Bzdak:2019pkr}. These different colliders have different stopping powers. The chemical potentials associated with each of them vary, as well. A precise estimation for the baryon chemical potential is crucial, especially for mapping out the QCD phase diagram \cite{Tawfik:2004vv,Tawfik:2004sw}. Ton this end, the conventional method is the one based on a statistical correspondence between multiplicities of the particles produced from high-energy collisions and thermodynamic quantities estimated in theoretical approaches \cite{Tawfik:2019yxn,Tawfik:2018sji}, such as the statistical thermal models \cite{Tawfik:2014eba}, namely the particle number and their ratios \cite{Tawfik:2016jzk,Tawfik:2013eua,Tawfik:2013dba,Tawfik:2012si,
Tawfik:2005qn,Tawfik:2004ss,Tawfik:2004vv}. Recently, an almost-entirely empirical estimation for the full chemical potential has been proposed \cite{Tawfik:2019gpc}. This is based on analyzing the transverse momentum distributions of the six well-identified produced particles, $\pi^+$, $\pi^-$, $K^+$, $K^-$, $p$, and $\bar{p}$. Deriving a generic expression for the dependence of the full chemical potential, the one combining various quantum numbers, on rapidity $y$, which is then utilized in reproducing the experimental results of the most central $p_{\perp}$ and $d^2 N/2 \pi p_{\bot} dp_{\bot} dy$, at $7.7$, $11.5$, $19.6$, $27$, $39$, $130$, $200~$GeV, a generic expression for the $y$-dependence of $\mu$ for the entire set of particle yields, at different energies, could be obtained; $\mu=a+b y^2$. We have also obtained an expression for the dependence of the collision energy on $\mu$, namely $\sqrt{s_{\mathtt{NN}}}=c[(\mu-a)/b]^{d/2}$. It was found that the proposed approach reproduces excellently the rapidity spectra of various particle yields measured, at different energies. 
 
The present script refines this procedure in the way that additional five particles are combined with, $K_s^0$, $\Lambda$, $\bar{\Lambda}$, $\Xi^-$, $\Xi^+$, i.e. more strangeness contents are added so that the share of the related chemical potential gains significance. We also cover more collision energies. Comparing with ref. \cite{Tawfik:2019gpc}, the present calculations improve the precision of the dependence of $\mu$ on $y$ and accordingly the entire approach, one one hand. One the other hand, it enables us to estimate various types of chemical potentials.

The present paper is organized as follows. The theoretical approach is discussed in section \ref{sec:app}. Section \ref{sec:res} gives details about the results obtained. Also, the dependence of the resulting chemical potential for strange and charged particles on the rapidity shall be presented in section \ref{sec:res}. Section \ref{sec:cncl} is devoted to our final conclusions.

\section{Theoretical Approach}
\label{sec:app}

Using the partition function of the hadron resonance gas model, the momentum distribution of a specific particle in the grand-canonical ensemble can be expressed as \cite{Tawfik:2014eba},
\begin{eqnarray}
E \frac{d^3 N}{d^3 p}\propto \left\lbrace \pm E \left(\exp\left[ \frac{E-\mu}{T}\right] \pm1\right) ^{-1} \right\rbrace, 
\label{eq:1}
\end{eqnarray}
and the volume element of momentum space reads
\begin{eqnarray}
\frac{d^3 p}{E} = m_{\mathrm{T}} d m_{\mathrm{T}} dy d \phi.
\label{eq:2}
\end{eqnarray}
Then the transverse momentum distribution $p_{\mathrm{T}}$ of the different particles emitted from the relativistic heavy-ion collisions can be expressed also by an exponential function,
\begin{eqnarray}
\frac{d^2 N}{2 \pi p_{\mathrm{T}} dp_{\mathrm{T}} dy}\propto \left\lbrace \pm m_{\mathrm{T}} \cosh(y)\left(\exp\left[ \frac{m_{\mathrm{T}}\cosh(y)-\mu}{T}\right] \pm1\right) ^{-1} \right\rbrace, 
\label{eq:3}
\end{eqnarray}
where $\pm$ stands for fermions and bosons, respectively. $E = m_{\mathrm{T}} \cosh(y)$, $m_{\mathrm{T}}$ is the transverse mass of the particle produced, which depends on both its transverse momentum $p_{\mathrm{T}}$ and its mass $m_{\mathrm{T}}=\sqrt{m^2+p_{\mathrm{T}}^2}$. Thus, the proportional constant reads, $C=\frac{g V}{(2 \pi)^3}$ with $g$ is the degeneracy factor and $V$ is the volume of the system
\begin{eqnarray}
\frac{d^2 N}{2 \pi p_{\mathrm{T}} dp_{\mathrm{T}} dy} &=& \pm \frac{g V}{(2 \pi)^3} m_{\mathrm{T}} \cosh(y)  \left(\exp\left[\frac{m_{\mathrm{T}} \cosh(y)-\mu}{T}\right] \pm 1 \right)^{-1}.
\label{eq:5}
\end{eqnarray}
The dependence of the chemical potential $\mu$ on the rapidity $y$ could be proposed as \cite{Tawfik:2019gpc},
\begin{eqnarray}
\mu = m_{\mathrm{T}} \cosh(y) - T \ln \left[\frac{\pm \frac{g V}{(2 \pi)^3} m_{\mathrm{T}} \cosh(y)}{\frac{d^2 N}{2 \pi p_{\mathrm{T}} dp_{\mathrm{T}} dy}} \mp 1\right],
\label{eq:6}
\end{eqnarray}
It should be noticed that $\mu$ sums up all types of chemical potentials related to the various quantum numbers, $\mu=n_B\mu_B+n_S\mu_S+n_Q\mu_Q+\cdots$. It is worth highlighting the difficulties of estimating these various types of chemical potentials. For a recent review, the readers are advised to consult ref. \cite{Tawfik:2018sji,Tawfik:2014dha,Tawfik:2014eba}. These are mainly constained by various laws conservation and strongly depending on various sophisticated observations. The present work aims at a precise estimation for the full $\mu$ as well as for $\mu_B$, $\mu_S$ and $\mu_Q$. This can be achieved through a precise estimation for the various components. Relative to ref. \cite{Tawfik:2019gpc}, we take into account here more strangeness contents; $K_s^0$, $\Lambda$, $\bar{\Lambda}$, $\Xi^-$, $\Xi^+$, which also brings more electric charges, and cover more collision energies, as well. 

\section{Results and discussion}
\label{sec:res}

%%%%%%%%%%%%%%%%%%%%%%%%%
\begin{figure}[!htb]
\includegraphics[width=5.cm]{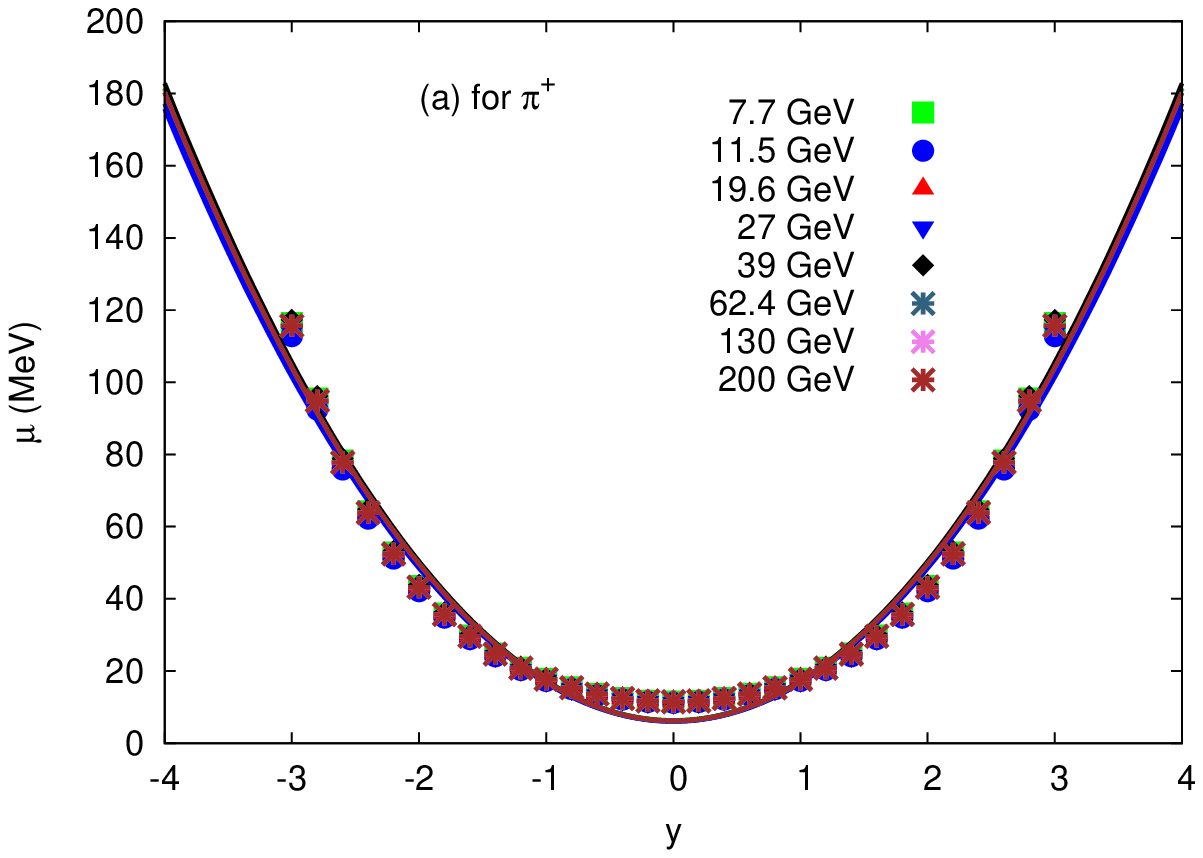}
\includegraphics[width=5.cm]{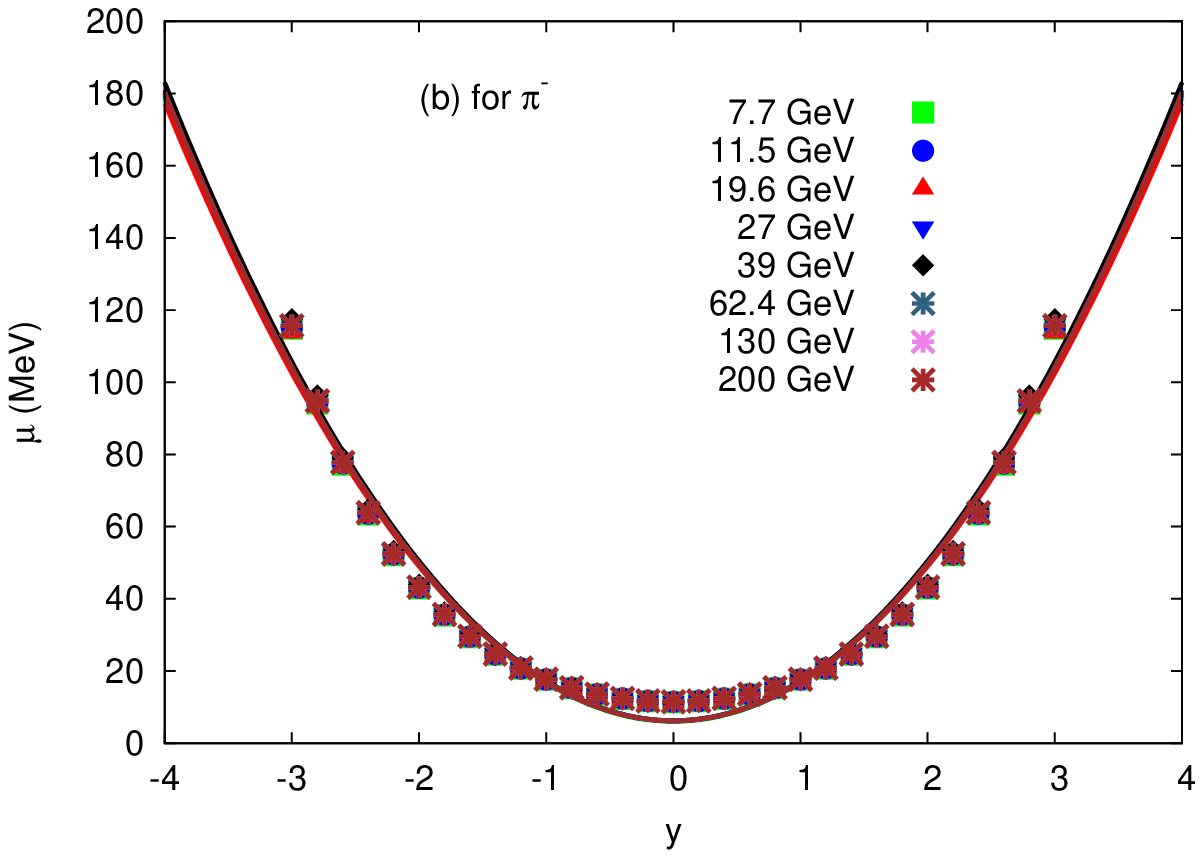}
\includegraphics[width=5.cm]{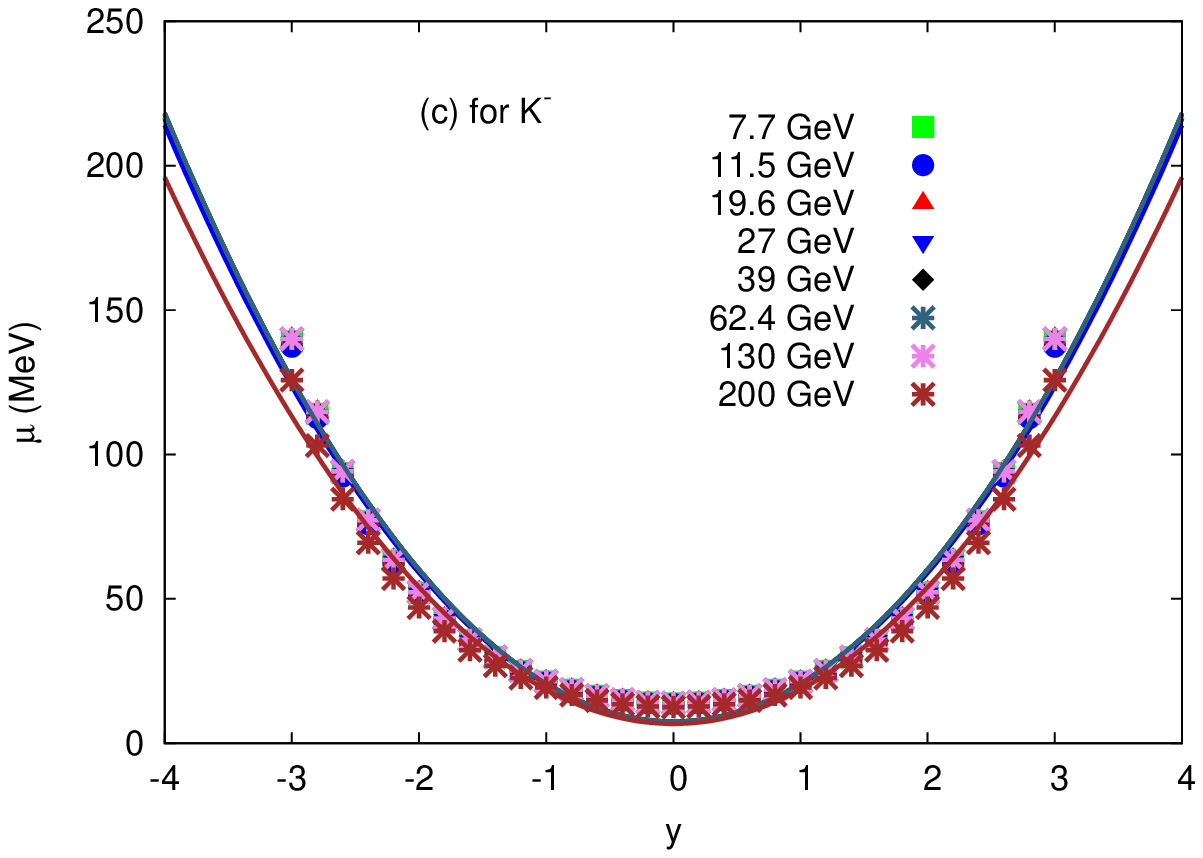}
\includegraphics[width=5.cm]{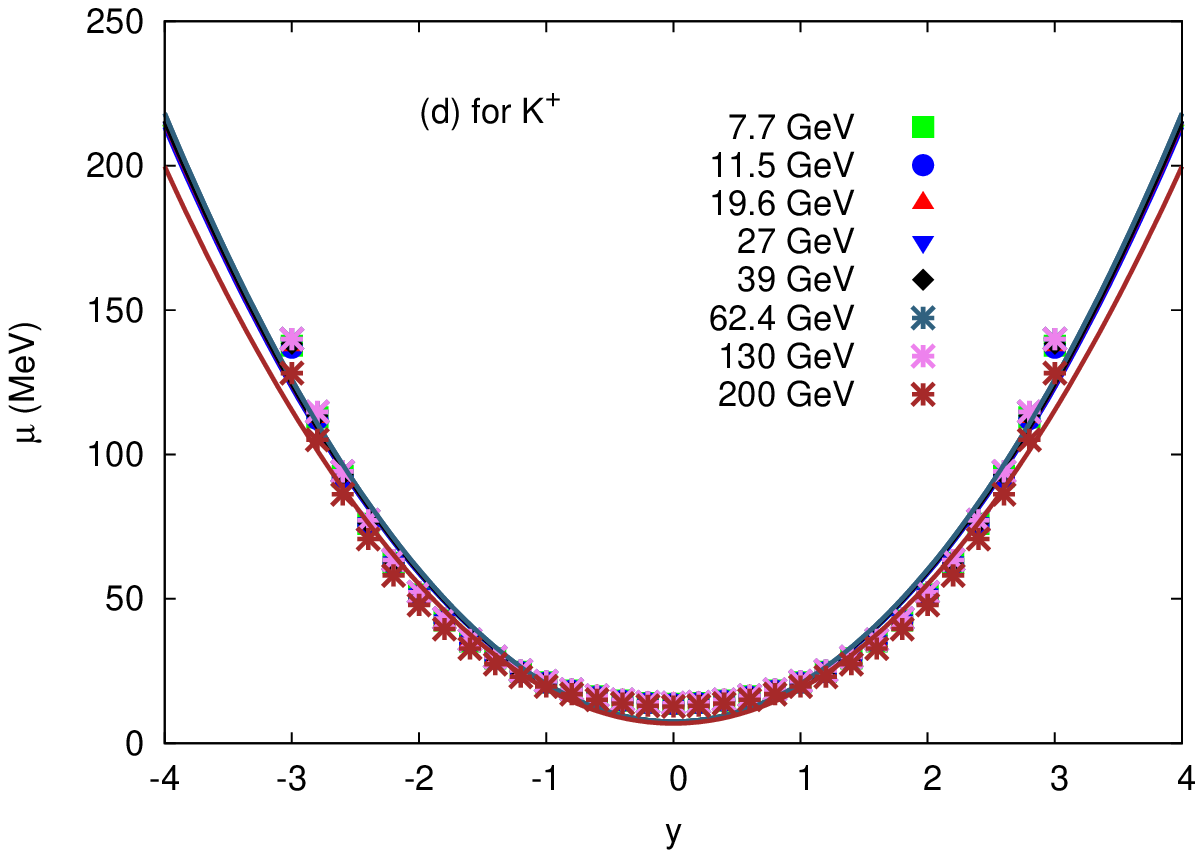} 
\includegraphics[width=5.cm]{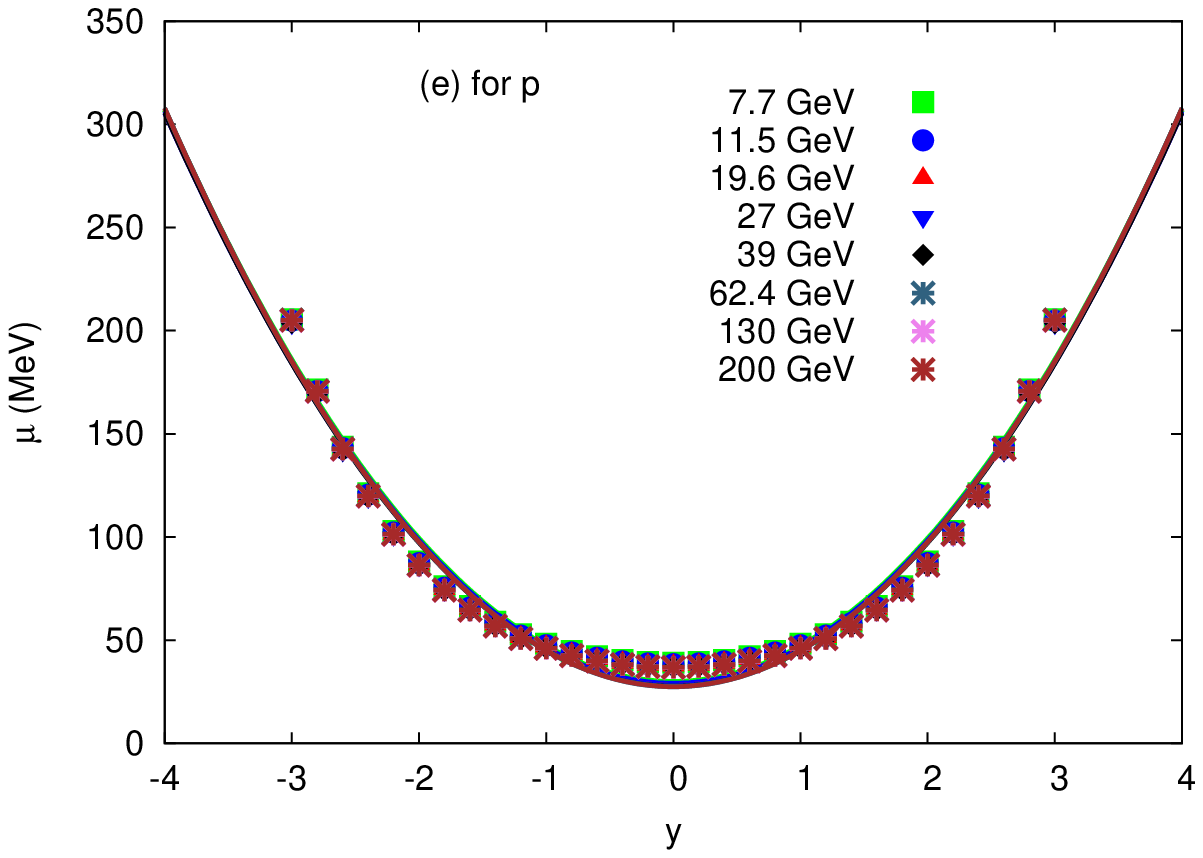}
\includegraphics[width=5.cm]{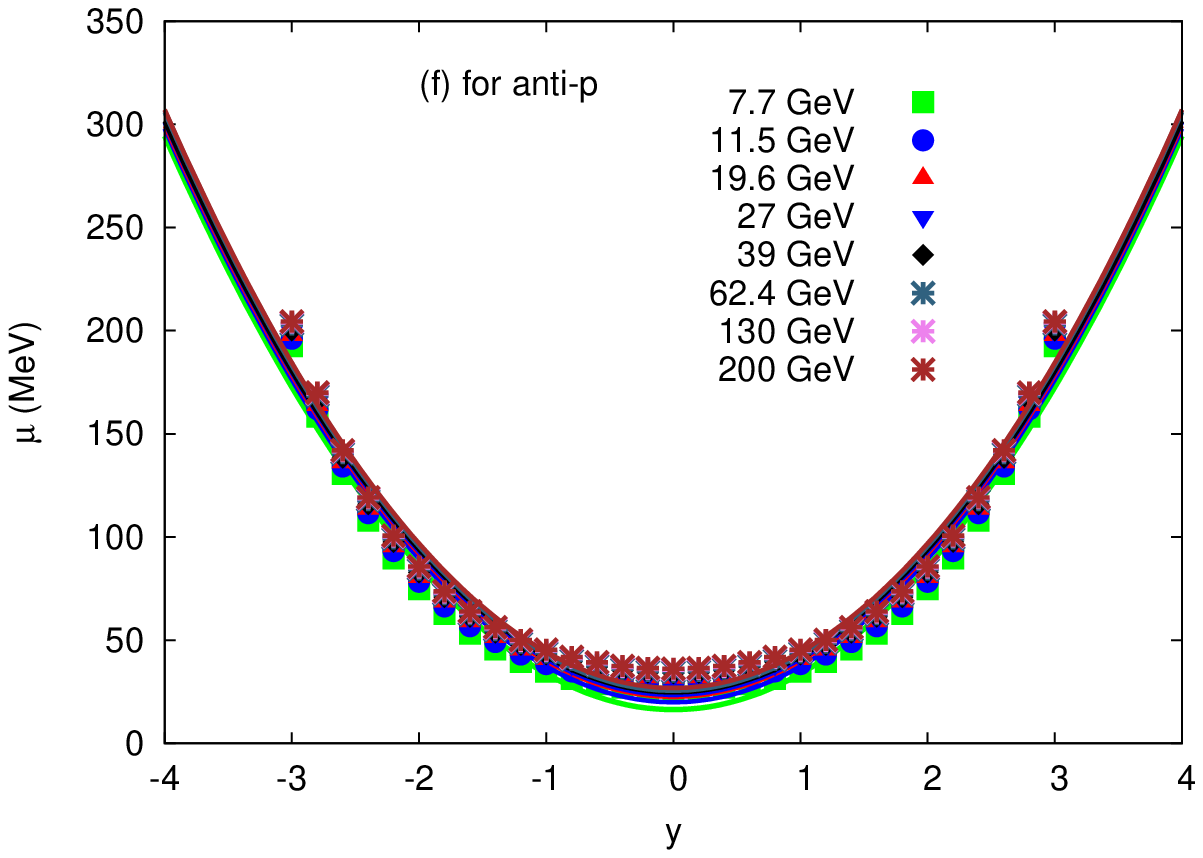} 
\caption{The full chemical potential $\mu$ as a function of the rapidity $y$ for $\pi^+$, $\pi^-$, $K^+$, $K^-$, $p$, $\bar{p}$ are depicted in panels (a), (b), (c), (d), (e), and (f), respectively. Symbols refer to the calculations based on Eq. (\ref{eq:6}), in which the STAR results on $p_{\mathrm{T}}$ in GeV-units and $d^2 N/(2 \pi p_{\mathrm{T}} dp_{\mathrm{T}} dy)$ in GeV$^{-2}$-units, at $\sqrt{s_{\mathtt{NN}}}=7.7$, $11.5$, $19.6$, $27$, $39$, $62.4$, $130$, $200~$GeV \cite{Adamczyk:2017iwn} are taken into account, while the curves represent the statistical fits, Tabs. \ref{Tab:1}-\ref{Tab:6} and \ref{Tab:12}.  \label{fig:1} }
\end{figure}
%%%%%%%%%%%%%%%%%%%%%%%%%%%

%%%%%%%%%%%%%%%%%%%%%%%%%
\begin{figure}[!htb]
\includegraphics[width=5.cm]{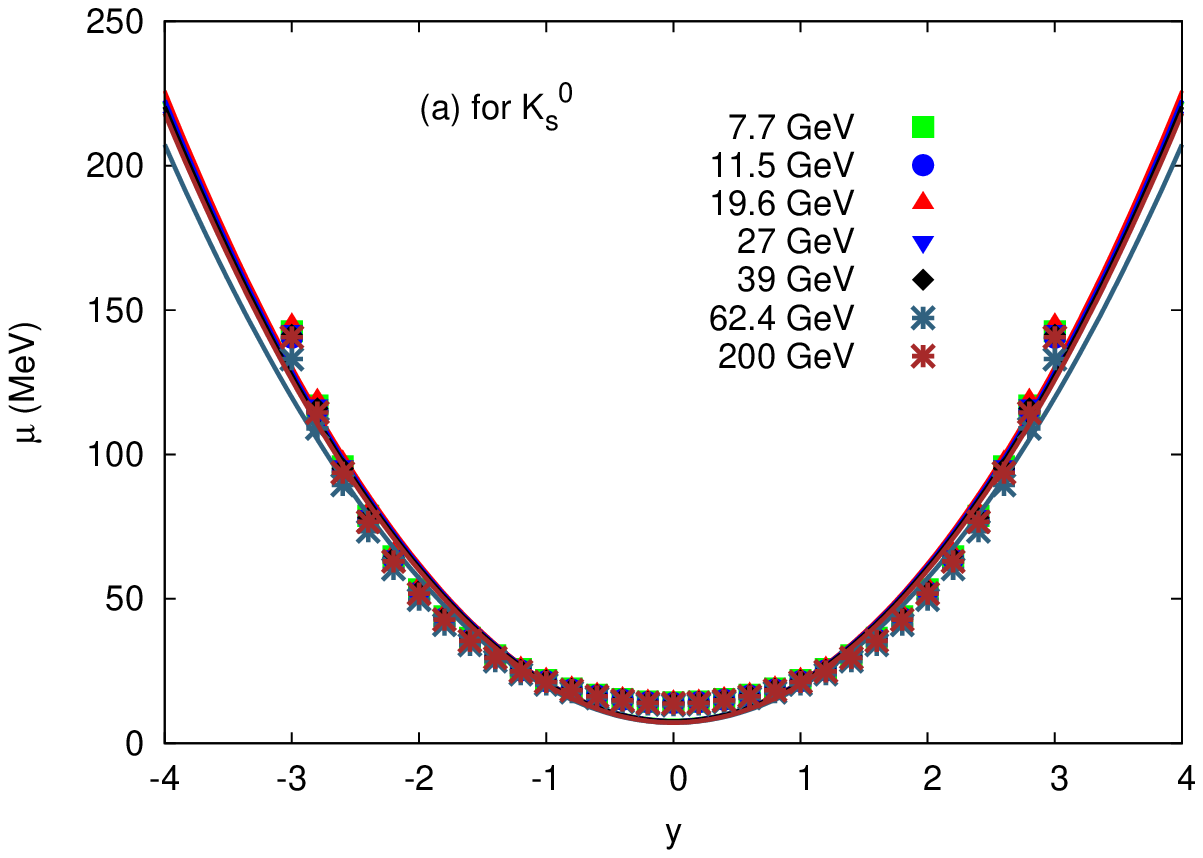}
\includegraphics[width=5.cm]{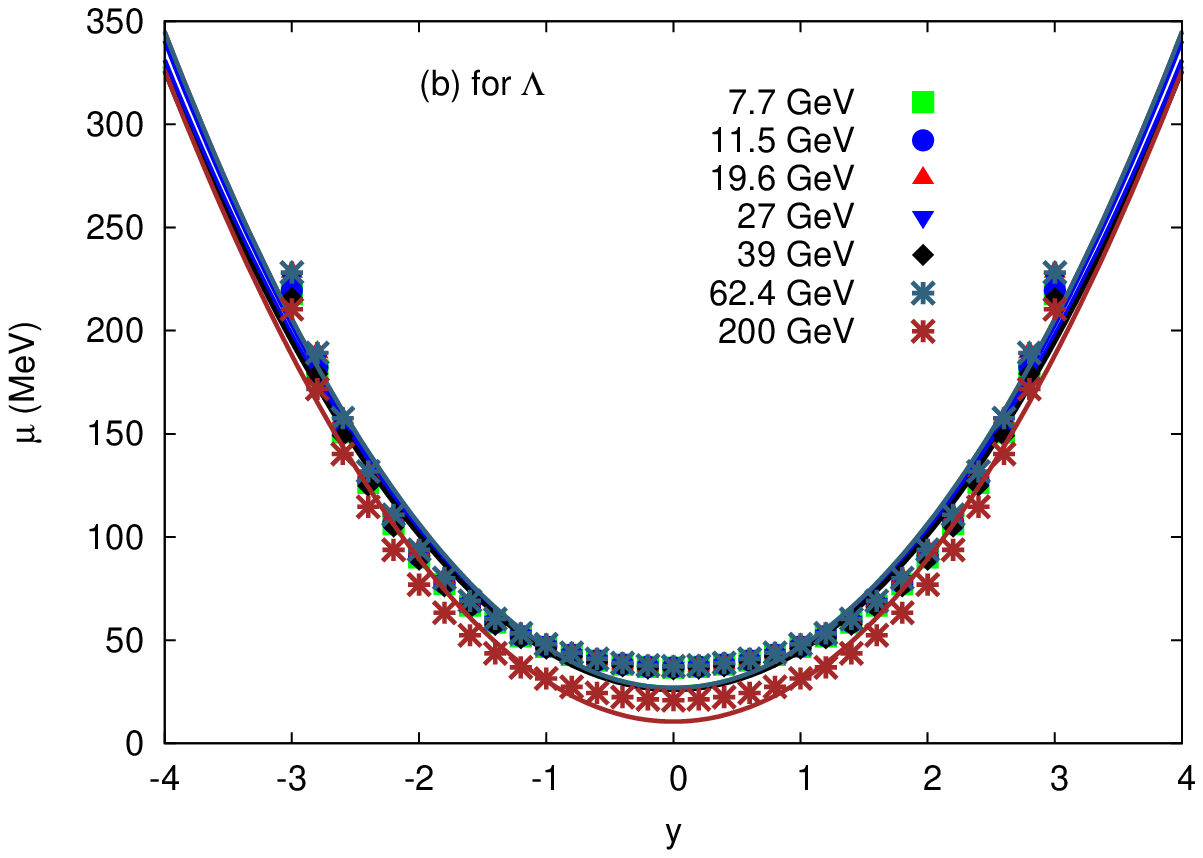} 
\includegraphics[width=5.cm]{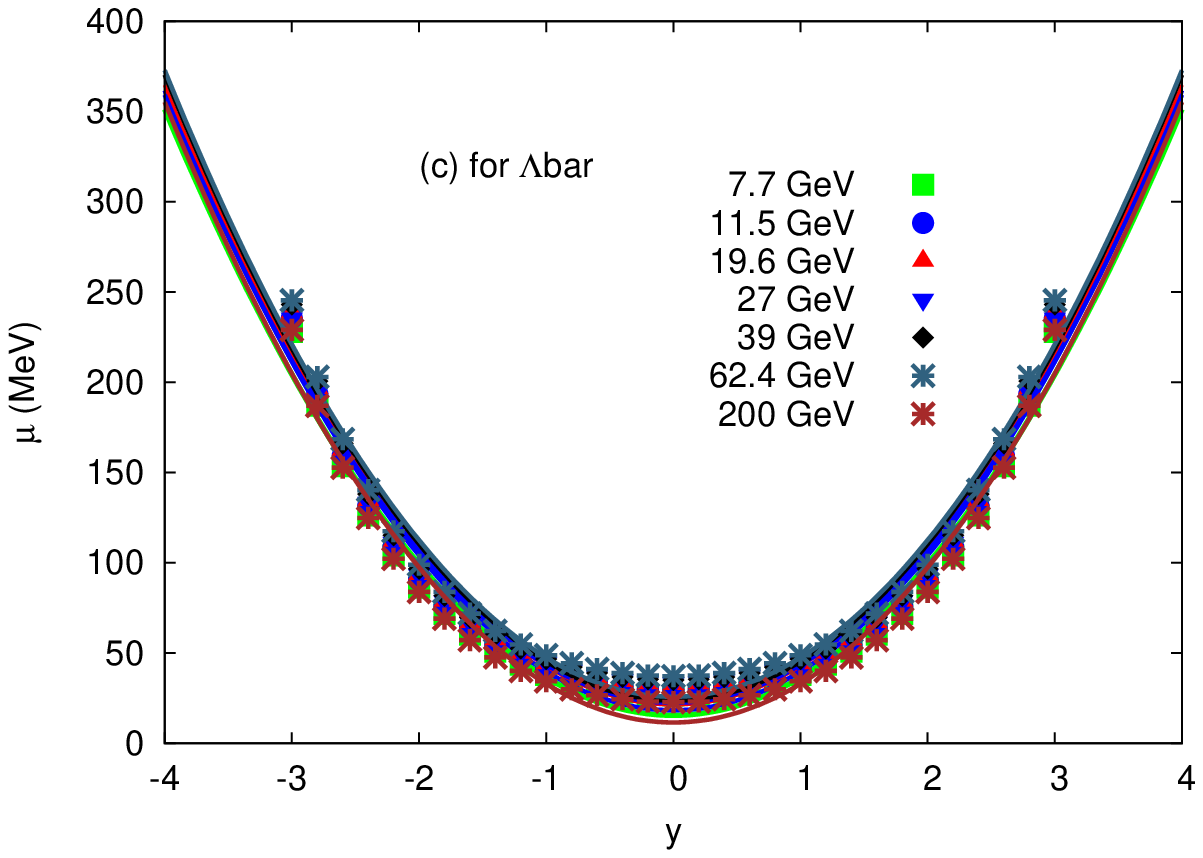}
\includegraphics[width=5.cm]{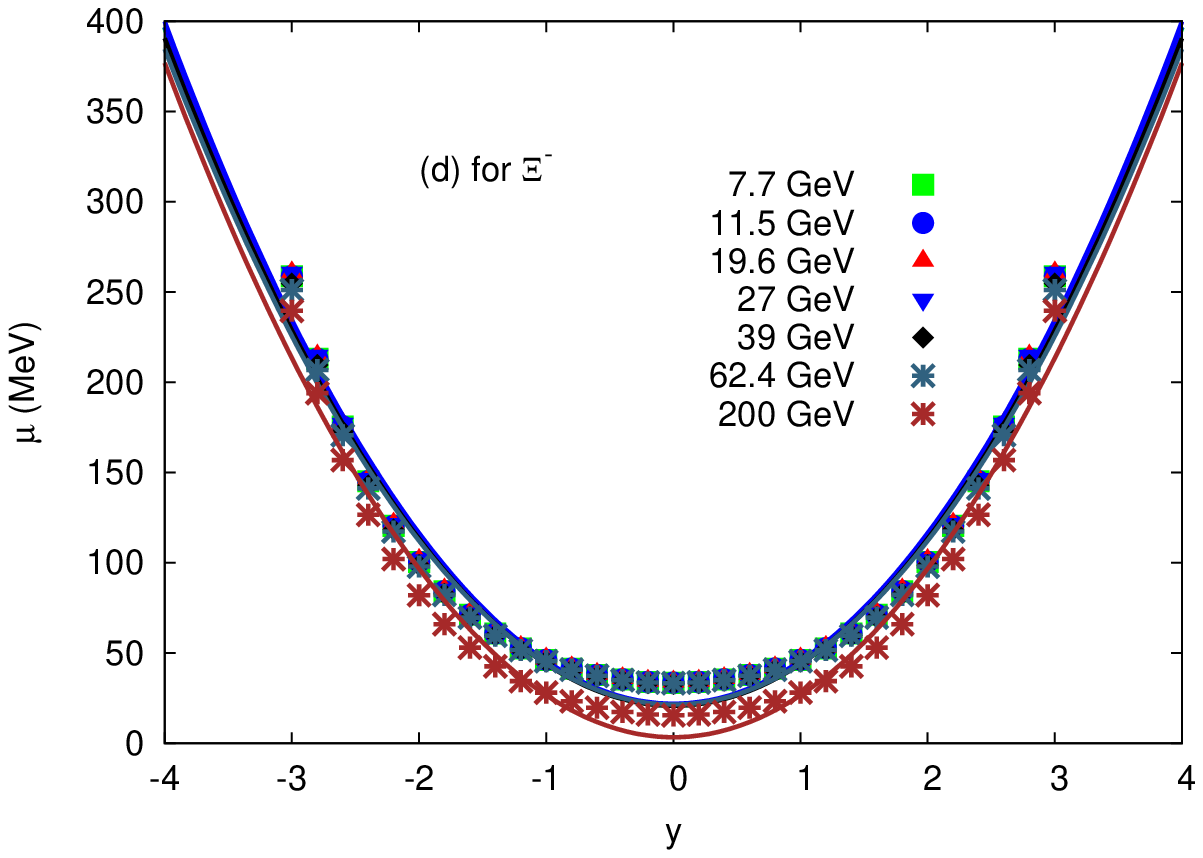} 
\includegraphics[width=5.cm]{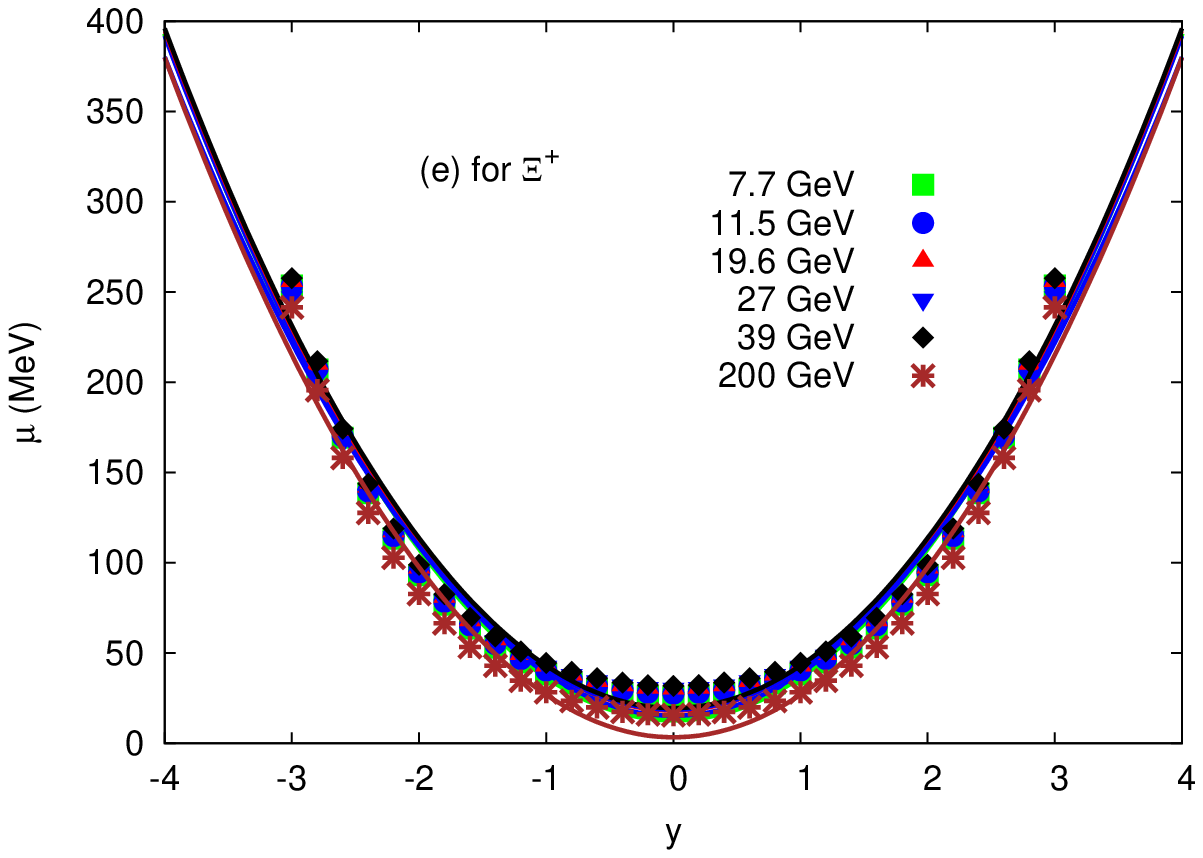}
\caption{The same as in Fig. \ref{fig:1} but here for $K^0_s$, $\Lambda$, $\bar{\Lambda}$, $\Xi^-$, $\Xi^+$, at $\sqrt{s_{\mathtt{NN}}}=7.7$, $11.5$, $19.6$, $27$, $39$, $62.4$, $200~$GeV \cite{Adam:2019koz,Aggarwal:2010ig,Abelev:2006cs}. The parameters deduced from the statistical fits are listed in Tabs. \ref{Tab:7}-\ref{Tab:12}. \label{fig:2}}
\end{figure}
%%%%%%%%%%%%%%%%%%%%%%%%%%%

In our previous work \cite{Tawfik:2019gpc}, we have proposed expressions for the dependence of the chemical potential on the rapidity for the well-identified particles, $\pi^+$, $\pi^-$, $K^+$, $K^-$, $p$, $\bar{p}$. In the present study, we cover more collision energies and take into consideration more strange particles, $K^0_s$, $\Lambda$, $\bar{\Lambda}$, $\Xi^-$, $\Xi^+$. The generic expression for the dependence of the chemical potential on the rapidity which was proposed in ref. \cite{Tawfik:2019gpc} is used here, as well, i.e. it seems to work perfectly for particles with considerable strangeness and electric charge contents, at energies ranging between $7.7$ and $200~$GeV. 

For our calculations, we use measurements for the transverse momentum $p_{\mathrm{T}}$ and the transverse momentum distribution $d^2 N/(2 \pi p_{\mathrm{T}} dp_{\mathrm{T}} dy)$ for each of the eleven particles, at certain collision energy. The results are depicted in Figs. \ref{fig:1}, \ref{fig:2} for $\pi^+$, $\pi^-$, $K^+$, $K^-$, $p$, $\bar{p}$ and $K^0_s$, $\Lambda$, $\bar{\Lambda}$, $\Xi^-$, $\Xi^+$, respectively. These calculations (symbols) have been fitted. The fit parameters are listen in Tabs. \ref{Tab:1}-\ref{Tab:11}. In these calculations, we assume that the freezeout temperature is $0.165~$GeV and keep this fixed for all particles, at the entire range of collision energies.

Figure \ref{fig:1} depicts the full chemical potential $\mu$ as a function of the rapidity $y$. Our calculations stemming from Eq. (\ref{eq:6}) are shown as symbols, where $p_{\mathrm{T}}$ in GeV-units and $d^2 N/(2 \pi p_{\mathrm{T}} dp_{\mathrm{T}} dy)$ in GeV$^{-2}$-units for $\pi^+$, $\pi^-$, $K^+$, $K^-$, $p$, $\bar{p}$, at $\sqrt{s_{\mathtt{NN}}}=7.7$, $11.5$, $19.6$, $27$, $39$, $130$, $200~$GeV are taken from ref. \cite{Adamczyk:2017iwn}. The resulting $\mu$ is then converted to MeV-units. These calculations (symbols) are fitted to the expressions outlined in Tab. \ref{Tab:12}. The resulting fit parameters are detailed in Tabs. \ref{Tab:1}-\ref{Tab:6}. The expressions proposed for the statistical fits are also depicted as curves. 

Figure \ref{fig:2} depicts the same as Fig. \ref{fig:1} but here for $K^0_s$, $\Lambda$, $\bar{\Lambda}$, $\Xi^-$, $\Xi^+$, at $\sqrt{s_{\mathtt{NN}}}=7.7$, $11.5$, $19.6$, $27$, $39$, $62.4$, $200~$GeV, where $p_{\mathrm{T}}$ in GeV-units and $d^2 N/(2 \pi p_{\mathrm{T}} dp_{\mathrm{T}} dy)$ in GeV$^{-2}$-units are taken from \cite{Adam:2019koz,Aggarwal:2010ig,Abelev:2006cs}. The resulting fit parameters are listed in Tabs. \ref{Tab:7}-\ref{Tab:11}. The expressions proposed for the statistical fits are given Tab. \ref{Tab:12}, as well. They are presented as curves.

\begin{table}[ht]
\centering % used for centering table
\begin{tabular}{||c | c | c | c | c ||} \hline\hline
  $\sqrt{s_{\mathtt{NN}}}$ (GeV) & $p_{\mathrm{T}}$ (GeV) & $\frac{d^2 N}{2 \pi p_{\mathrm{T}} dp_{\mathrm{T}} dy}~$(GeV)$^{-2}$ &  $a$ (MeV) & $b$ (MeV)\\
   \hline
  7.7 & 0.68 & 12.023 & $6.217\pm1.418$ & $10.949\pm0.331$\\
  11.5 & 0.657 & 15.8111 & $6.021\pm1.373$ & $10.604\pm0.32$\\
  19.6 & 0.68 & 21.7961 & $6.221\pm1.419$ & $10.955\pm0.331$\\
  27 & 0.663 & 24.669 & $6.073\pm1.385$ & $10.694\pm0.323$\\
  39 & 0.685 & 25.724 & $6.264\pm1.429$ & $11.032\pm0.333$ \\
  62.4 & 0.675 & 30.9 & $6.177\pm1.409$ & $10.878\pm0.328$ \\
  130 & 0.675 & 63.9 & $6.177\pm1.409$ & $10.877\pm0.328$ \\
  200 & 0.675 & 44.3 & $6.177\pm1.409$ & $10.878\pm0.328$ \\
  \hline\hline 
  \end{tabular}
\caption{For $\pi^+$, the experimental values for $p_{\mathrm{T}}$ in GeV-units and $d^2 N/(2 \pi p_{\mathrm{T}} dp_{\mathrm{T}} dy)$ in GeV$^{-2}$-units, at various energies substitueted in Eq. (\ref{eq:6}) in order to estimate the full chemical potential depicted in Fig. \ref{fig:1} are taken from \cite{Adamczyk:2017iwn}. The fit parameters, namely $a$ and $b$, are obtained, at $m_{\pi^+}=0.140~$GeV and $g_{\pi^+}=1$.  \label{Tab:1} }
\end{table}
%%%%%%%%%%%%%%%%%%%%%%
%

\begin{table}[ht]
\centering % used for centering table
\begin{tabular}{||c | c | c | c | c ||} \hline\hline
  $\sqrt{s_{\mathtt{NN}}}$ (GeV) & $p_{\mathrm{T}}$ (GeV) & $\frac{d^2 N}{2 \pi p_{\mathrm{T}} dp_{\mathrm{T}} dy}~$(GeV)$^{-2}$ &  $a$ (MeV) & $b$ (MeV)\\
   \hline
  7.7 & 0.67 & 11.403 & $6.131\pm1.398$ & $10.798\pm0.326$\\
  11.5 & 0.673 & 16.669 & $6.155\pm1.404$ & $10.84\pm0.327$\\
  19.6 & 0.663 & 21.236 & $6.069\pm1.384$ & $10.688\pm0.323$\\
  27 & 0.678 & 23.792 & $6.2\pm1.414$ & $10.919\pm0.33$\\
  39 & 0.686 & 25.292 & $6.273\pm1.431$ & $11.048\pm0.334$ \\
  62.4 & 0.675 & 31.4 & $6.177\pm1.409$ & $10.878\pm0.328$ \\
  130 & 0.675 & 37.6 & $6.177\pm1.409$ & $10.877\pm0.328$ \\
  200 & 0.675 & 44.8 & $6.177\pm1.409$ & $10.877\pm0.328$ \\
  \hline\hline 
  \end{tabular}
\caption{The same as Tab. \ref{Tab:1} but for $m_{\pi^-}=0.140~$GeV and $g_{\pi^-}=1$.  \label{Tab:2} }
\end{table}
%%%%%%%%%%%%%%%%%%%%%%%
%

\begin{table}[ht]
\centering % used for centering table
\begin{tabular}{||c | c | c | c | c ||} \hline\hline
  $\sqrt{s_{\mathtt{NN}}}$ (GeV) & $p_{\mathrm{T}}$ (GeV) & $\frac{d^2 N}{2 \pi p_{\mathrm{T}} dp_{\mathrm{T}} dy}~$(GeV)$^{-2}$ &  $a$ (MeV) & $b$ (MeV)\\
   \hline
  7.7 & 0.669 & 1.779 & $7.433\pm1.695$ & $13.09\pm0.395$\\
  11.5 & 0.655 & 2.672 & $7.332\pm1.672$ & $12.913\pm0.39$\\
  19.6 & 0.676 & 4.264 & $7.484\pm1.707$ & $13.179\pm0.398$\\
  27 & 0.666 & 5.747 & $7.411\pm1.69$ & $13.052\pm0.394$\\
  39 & 0.674 & 5.982 & $7.465\pm1.702$ & $13.146\pm0.397$ \\
  62.4 & 0.675 & 7.73 & $7.474\pm1.704$ & $13.162\pm0.397$ \\
  130 & 0.675 & 8.86 & $7.474\pm1.704$ & $13.162\pm0.397$ \\
  200 & 0.567 & 11.711 & $6.714\pm1.531$ & $11.824\pm0.357$ \\
  \hline\hline 
  \end{tabular}
\caption{The same as Tab. \ref{Tab:1} but for $m_{K^-}=0.490~$GeV and $g_{K^-}=1$.  \label{Tab:3} }
\end{table}
%%%%%%%%%%%%%%%%%%%
%

\begin{table}[ht]
\centering % used for centering table
\begin{tabular}{||c | c | c | c | c ||} \hline\hline
  $\sqrt{s_{\mathtt{NN}}}$ (GeV) & $p_{\mathrm{T}}$ (GeV) & $\frac{d^2 N}{2 \pi p_{\mathrm{T}} dp_{\mathrm{T}} dy}~$(GeV)$^{-2}$ &  $a$ (MeV) & $b$ (MeV)\\
   \hline
  7.7 & 0.658 & 5.014 & $7.349\pm1.671$ & $12.942\pm0.391$\\
  11.5 & 0.653 & 5.464 & $7.313\pm1.668$ & $12.878\pm0.389$\\
  19.6 & 0.658 & 6.552 & $7.349\pm1.676$ & $12.942\pm0.391$\\
  27 & 0.662 & 7.537 & $7.379\pm1.683$ & $12.994\pm0.392$\\
  39 & 0.663 & 6.888 & $7.386\pm1.684$ & $13.006\pm0.393$ \\
  62.4 & 0.675 & 8.71 & $7.473\pm1.704$ & $13.16\pm0.397$ \\
  130 & 0.675 & 9.78 & $7.474\pm1.704$ & $13.162\pm0.397$ \\
  200 & 0.586 & 11.912 & $6.844\pm1.561$ & $12.052\pm0.364$ \\
  \hline\hline 
  \end{tabular}
\caption{The same as Tab. \ref{Tab:1} but for $m_{K^+}=0.490~$GeV and $g_{K^+}=1$.  \label{Tab:4} }
\end{table}
%%%%%%%%%%%%%%%%%%
%

\begin{table}[ht]
\centering % used for centering table
\begin{tabular}{||c | c | c | c | c ||} \hline\hline
  $\sqrt{s_{\mathtt{NN}}}$ (GeV) & $p_{\mathrm{T}}$ (GeV) & $\frac{d^2 N}{2 \pi p_{\mathrm{T}} dp_{\mathrm{T}} dy}~$(GeV)$^{-2}$ &  $a$ (MeV) & $b$ (MeV)\\
   \hline
  7.7 & 0.653 & 13.335 & $29.823\pm2.414$ & $17.322\pm0.563$\\
  11.5 & 0.653 & 10.467 & $29.157\pm2.414$ & $17.322\pm0.563$\\
  19.6 & 0.651 & 7.796 & $28.338\pm2.411$ & $17.3\pm0.562$\\
  27 & 0.653 & 7.389 & $28.199\pm2.414$ & $17.321\pm0.563$\\
  39 & 0.659 & 5.995 & $27.643\pm2.42$ & $17.369\pm0.564$ \\
  62.4 & 0.675 & 5.84 & $27.634\pm2.439$ & $17.52\pm0.569$ \\
  130 & 0.675 & 5.18 & $27.305\pm2.439$ & $17.52\pm0.569$ \\
  200 & 0.675 & 5.44 & $27.439\pm2.439$ & $17.52\pm0.569$ \\
  \hline\hline 
  \end{tabular}
\caption{The same as Tab. \ref{Tab:1} but for $m_{p}=0.938~$GeV and $g_{p}=2$.  \label{Tab:5} }
\end{table}
%%%%%%%%%%%%%%%%%%%%%%%  
%

\begin{table}[ht]
\centering % used for centering table
\begin{tabular}{||c | c | c | c | c ||} \hline\hline
  $\sqrt{s_{\mathtt{NN}}}$ (GeV) & $p_{\mathrm{T}}$ (GeV) & $\frac{d^2 N}{2 \pi p_{\mathrm{T}} dp_{\mathrm{T}} dy}~$(GeV)$^{-2}$ &  $a$ (MeV) & $b$ (MeV)\\
   \hline
  7.7 & 0.659 & 0.098 & $16.32\pm2.42$ & $17.371\pm0.564$\\
  11.5 & 0.659 & 0.357 & $19.892\pm2.421$ & $17.375\pm0.564$\\
  19.6 & 0.651 & 0.864 & $22.288\pm2.411$ & $17.3\pm0.562$\\
  27 & 0.656 & 1.307 & $23.446\pm2.417$ & $17.348\pm0.563$\\
  39 & 0.651 & 2.057 & $24.675\pm2.411$ & $17.304\pm0.562$ \\
  62.4 & 0.675 & 2.63 & $25.441\pm2.439$ & $ 17.52\pm0.569$ \\
  130 & 0.675 & 3.67 & $26.357\pm2.439$ & $17.52\pm0.569$ \\
  200 & 0.675 & 4.29 & $26.786\pm2.439$ & $17.52\pm0.569$ \\
  \hline\hline 
  \end{tabular}
\caption{The same as Tab. \ref{Tab:1} but for $m_{\bar{p}}=0.938~$GeV and $g_{\bar{p}}=2$.  \label{Tab:6} }
\end{table}
%%%%%%%%%%%%%%%
%

\begin{table}[ht]
\centering % used for centering table
\begin{tabular}{||c | c | c | c | c ||} \hline\hline
  $\sqrt{s_{\mathtt{NN}}}$ (GeV) & $p_{\mathrm{T}}$ (GeV) & $\frac{d^2 N}{2 \pi p_{\mathrm{T}} dp_{\mathrm{T}} dy}~$(GeV)$^{-2}$ &  $a$ (MeV) & $b$ (MeV) \\
   \hline
  7.7 & 0.692 & 2.576 & $7.617\pm1.737$ & $13.414\pm0.405$ \\
  11.5 & 0.676 & 3.679 & $7.498\pm1.71$ & $13.205\pm0.399$ \\
  19.6 & 0.709 & 4.693 & $7.74\pm1.765$ & $13.631\pm0.411$ \\
  27 & 0.694 & 6.032 & $7.628\pm1.74$ & $13.434\pm0.406$ \\
  39 & 0.683 & 5.603 & $7.553\pm1.722$ & $13.301\pm0.402$ \\
  62.4 & 0.62 & 6.918 & $7.103\pm1.62$ & $12.509\pm0.378$ \\
  200 & 0.643 & 0.041 & $7.156\pm1.771$ & $13.182\pm0.413$ \\
  \hline\hline 
  \end{tabular}
\caption{The same as Tab. \ref{Tab:1} but for $K_s^0$, where $p_{\mathrm{T}}$ in GeV-units and $d^2 N/(2 \pi p_{\mathrm{T}} dp_{\mathrm{T}} dy)$ in GeV$^{-2}$-units are taken from \cite{Adam:2019koz,Aggarwal:2010ig,Abelev:2006cs}. For the resulting fit parameters, we use $m_{K_s^0}=0.4937~$GeV and $g_{K_s^0}=1$.  \label{Tab:7} }
\end{table}
%%%%%%%%%%%%%%%%%%%%%%%%%%
%

\begin{table}[ht]
\centering % used for centering table
\begin{tabular}{||c | c | c | c | c ||} \hline\hline
  $\sqrt{s_{\mathtt{NN}}}$ (GeV) & $p_{\mathrm{T}}$ (GeV) & $\frac{d^2 N}{2 \pi p_{\mathrm{T}} dp_{\mathrm{T}} dy}~$(GeV)$^{-2}$ &  $a$ (MeV) & $b$ (MeV)\\
   \hline
  7.7 & 0.541 & 3.398 & $26.706\pm2.611$ & $18.848\pm0.609$\\
  11.5 & 0.567 & 3.266 & $26.674\pm2.635$ & $19.03\pm0.614$\\
  19.6 & 0.682 & 2.767 & $26.601\pm2.749$ & $19.915\pm0.641$\\
  27 & 0.649 & 2.395 & $26.088\pm2.715$ & $19.649\pm0.633$\\
  39 & 0.525 & 2.61 & $25.934\pm2.597$ & $18.739\pm0.605$ \\
  62.4 & 0.677 & 3.02 & $26.823\pm2.744$ & $19.872\pm0.64$ \\
  200 & 0.659 & 0.008 & $10.451\pm2.726$ & $19.734\pm0.635$ \\
   \hline\hline 
  \end{tabular}
\caption{The same as Tab. \ref{Tab:7} but for $m_{\Lambda}=1.1157~$GeV and $g_{\Lambda}=2$ is presented. \label{Tab:8} }
\end{table}
%%%%%%%%%%%%%%%%%%%%%%%%%
%

\begin{table}[ht]
\centering % used for centering table
\begin{tabular}{||c | c | c | c | c ||} \hline\hline
  $\sqrt{s_{\mathtt{NN}}}$ (GeV) & $p_{\mathrm{T}}$ (GeV) & $\frac{d^2 N}{2 \pi p_{\mathrm{T}} dp_{\mathrm{T}} dy}~$(GeV)$^{-2}$ &  $a$ (MeV) & $b$ (MeV)\\
   \hline
  7.7 & 0.611 & 0.041 & $15.49\pm2.888$ & $20.981\pm0.673$\\
  11.5 & 0.676 & 0.105 & $18.291\pm2.949$ & $21.456\pm0.687$\\
  19.6 & 0.676 & 0.316 & $21.313\pm2.949$ & $21.456\pm0.687$\\
  27 & 0.621 & 0.493 & $22.359\pm2.897$ & $21.055\pm0.675$\\
  39 & 0.703 & 0.719 & $23.665\pm2.976$ & $21.664\pm0.694$ \\
  62.4 & 0.706 & 1.531 & $25.754\pm2.979$ & $21.689\pm0.694$ \\
  200 & 0.679 & 0.009 & $11.497\pm2.952$ & $21.478\pm0.688$ \\
  \hline\hline 
  \end{tabular}
\caption{The same as Tab. \ref{Tab:7} but for $m_{\bar{\Lambda}}=1.1157~$GeV and $g_{\bar{\Lambda}}=2$ is presented. \label{Tab:9} }
\end{table}
%%%%%%%%%%%%%%%%%%%%%%%
%

\begin{table}[ht]
\centering % used for centering table
\begin{tabular}{||c | c | c | c | c ||} \hline\hline
  $\sqrt{s_{\mathtt{NN}}}$ (GeV) & $p_{\mathrm{T}}$ (GeV) & $\frac{d^2 N}{2 \pi p_{\mathrm{T}} dp_{\mathrm{T}} dy}~$(GeV)$^{-2}$ &  $a$ (MeV) & $b$ (MeV)\\
   \hline
  7.7 & 0.775 & 0.223 & $21.247\pm3.209$ & $23.461\pm0.748$\\
  11.5 & 0.775 & 0.23 & $21.38 \pm3.208$ & $23.458\pm0.748$\\
  19.6 & 0.794 & 0.254 & $21.678\pm3.229$ & $23.615\pm0.753$\\
  27 & 0.794 & 0.276 & $21.91\pm3.229$ & $23.616\pm0.753$\\
  39 & 0.729 & 0.199 & $20.781\pm3.163$ & $23.105\pm0.737$ \\
  62.4 & 0.674 & 0.266 & $21.388 \pm3.11$ & $22.696\pm0.725$ \\
  200 & 0.76 & 0.0003 & $3.3294\pm3.193$ & $23.343\pm0.744$ \\
   \hline\hline 
  \end{tabular}
\caption{The same as Tab. \ref{Tab:7} but for $m_{\Xi^-}=1.3217~$GeV and $g_{\Xi^-}=2$ is presented.  \label{Tab:10} }
\end{table}
%%%%%%%%%%%%%%%%%%%%%%%
%
\begin{table}[ht]
\centering % used for centering table
\begin{tabular}{||c | c | c | c | c ||} \hline\hline
  $\sqrt{s_{\mathtt{NN}}}$ (GeV) & $p_{\mathrm{T}}$ (GeV) & $\frac{d^2 N}{2 \pi p_{\mathrm{T}} dp_{\mathrm{T}} dy}~$(GeV)$^{-2}$ &  $a$ (MeV) & $b$ (MeV)\\
   \hline
  7.7 & 0.813 & 0.01 & $12.967\pm3.248$ & $23.766\pm0.757$\\
  11.5 & 0.784 & 0.025 & $15.28\pm3.218$ & $23.531\pm0.75$\\
  19.6 & 0.784 & 0.063 & $17.813\pm3.218$ & $23.531\pm0.75$\\
  27 & 0.66 & 0.1 & $18.655\pm3.096$ & $22.592\pm0.722$\\
  39 & 0.784 & 0.117 & $19.051\pm3.218$ & $23.531\pm0.75$ \\
  200 & 0.785 & 0.0003 & $3.308\pm3.219$ & $23.538\pm0.75$ \\
  \hline\hline 
  \end{tabular}
\caption{The same as Tab. \ref{Tab:7} but for $m_{\Xi^+}=1.3217~$GeV and $g_{\Xi^+}=2$ is presented.  \label{Tab:11} }
\end{table}
%%%%%%%%%%%%%%%%%%%%%%%%%%%

In order to find out how the chemical potential $\mu$ depends on the rapidity $y$, we go as follows. First, we substitute with the experimental values of $p_{\mathrm{T}}~$GeV and $d^2 N/(2 \pi p_{\mathrm{T}} dp_{\mathrm{T}} dy)~$GeV$^{-2}$ measured in the most-central collisions \cite{Adamczyk:2017iwn,Abelev:2008ab,Adam:2019koz,Aggarwal:2010ig,Abelev:2006cs}, Tabs. \ref{Tab:1}-\ref{Tab:11}, in Eq. (\ref{eq:6}). Second, we draw $\mu$ vs. $y$ as shown in Figs. \ref{fig:1} and \ref{fig:2}. Third, we then propose analytical expressions for each of the produced particles with unknown variables $a$ and $b$ which are given in Tabs. \ref{Tab:1}-\ref{Tab:11}. Fourth, we conclude a general expression for each pair of particle and anti-particle.

For each of the particle pairs, we find that the proposed expression is nearly independent on the collision energies. The various expressions of each particle's chemical potential on rapidity are listed out in Tab. \ref{Tab:12}, i.e. for the eleven particles there are eleven expressions. The generic expression for all particles is the one which was given in ref. \cite{Tawfik:2019gpc}, namely
\begin{equation}
\mu = a + b y^2, \label{eq:muy}
\end{equation}
where precise estimations for the parameters $a$ and $b$ are given in Tab. \ref{Tab:12}.

\begin{table}[ht]
\centering % used for centering table
\begin{tabular}{|c|c|} \hline\hline
  Particle & $\mu (MeV) = a (MeV) + b (MeV) y^2$\\
   \hline
 $\pi^+$ & $\mu=(6.073\pm1.385) + (10.694\pm0.323)\, y^2$ \\ \hline 
 $\pi^-$ & $\mu=(6.069\pm1.384) + (10.688\pm0.323)\, y^2$ \\ \hline 
 $K^-$ & $\mu=(7.411\pm1.69) + (13.052\pm0.394)\, y^2$ \\ \hline 
 $K^+$ & $\mu=(7.349\pm1.676) + (12.942\pm0.391)\, y^2$ \\ \hline 
 $p$ & $\mu=(27.643\pm2.42) + (17.369\pm0.564)\, y^2$ \\ \hline 
 $\bar{p}$ & $\mu=(23.446\pm2.417) + (17.348\pm0.563)\, y^2$ \\ \hline 
 $K_s^0$ & $\mu=(7.498\pm1.71) + (13.205\pm0.399)\, y^2$ \\ \hline 
 $\Lambda$ & $\mu=(26.601\pm2.749) + (19.915\pm0.641)\, y^2$ \\ \hline 
 $\bar{\Lambda}$ & $\mu=(21.313\pm2.949) + (21.456\pm0.687)\, y^2$ \\ \hline 
 $\Xi^-$ & $\mu=(21.678\pm3.229) + (23.615\pm0.753)\, y^2$ \\ \hline 
 $\Xi^+$ & $\mu=(17.813\pm3.218) + (23.531\pm0.75)\, y^2$ \\
 \hline
 \end{tabular}
\caption{The parameters $a$ and $b$, Eq. (\ref{eq:muy}), for each of the eleven particles. \label{Tab:12} }
\end{table}
%%%%%%%%%%%%%%%%%%%%%%%%%%%

To judge about the goodness of the proposed expressions for the eleven particles, Tab. \ref{Tab:12}), we draw the rapidity vs. the center-of-mass energies for each particle (not shown here). Then, we fit these dependences. We get
\bea
\sqrt{s_{\mathtt{NN}}} [\mathtt{GeV}]= c [\mathtt{GeV}]\,  y^d,  \label{eq:sqrtsy}
\eea
where $c$ and $d$ are constants depending on the type of the particle. \begin{itemize}
\item For $\pi^+$:
$c              = 397.87          \pm 26.38$ GeV, %        (6.631%)
$d              = -2.173         \pm 0.136$, %       (6.254%)
\item For $\pi^-$:
$c               = 398.282         \pm 26.32$ GeV, %        (6.61%)
$d              = -2.176        \pm 0.135$, %       (6.228%)
\item For $K^-$:
$c               = 339.643         \pm 39.71$ GeV, %        (11.69%)
$d              = -2.245         \pm 0.284$, %       (12.64%)
\item For $K^+$:
$c               = 332.27          \pm 34.2$ GeV, %         (10.29%)
$d              = -2.220        \pm 0.253$, %       (11.4%)
\item For $p$:
$c               = 190.2         \pm 11.94$ GeV, %        (19.57%)
$d              = -2.2        \pm 0.079$, %      (18.18%)
\item For $\bar{p}$:
$c               = 180.054         \pm 8.858$ GeV, %        (12.43%)
$d              = -2.13       \pm 0.088$, %      (13.05%)
\item For $K_s^0$:
$c               = 318.825         \pm 9.915$ GeV, %        (3.11%)
$d               = -2.176         \pm 0.079$, %      (3.615%)
\item For $\Lambda$:
$c               = 166.698         \pm 6.922$ GeV, %        (4.153%)
$d               = -2.252        \pm 0.189$, %       (8.405%)
\item For $\bar{\Lambda}$:
$c               = 144.401         \pm 4.959$ GeV, %        (3.434%)
$d               = -2.125         \pm 0.141$, %       (6.643%)
\item For $\Xi^-$:
$c               = 181.8          \pm 17.66$ GeV, %        (10.2%)
$d               = -2.3         \pm 0.792$, %       (22.65%)
\item For $\Xi^+$:
$c               = 206.395          \pm 1.885$ GeV, %        (0.9131%)
$d               = -2.364         \pm 0.043$. %      (1.814%)
\end{itemize}
Then, when substituting Eq. (\ref{eq:muy}) into Eq. (\ref{eq:sqrtsy}), we get an expression for the dependence of the rapidity on the center-of-mass energies \cite{Tawfik:2019gpc}
\begin{eqnarray}
\sqrt{s_{\mathtt{NN}}}~\mathtt{[\mathtt{GeV}]} &=& c [\mathtt{GeV}]\left(\frac{\mu~\mathtt{[\mathtt{MeV}]}-a[\mathtt{MeV}]}{b[\mathtt{MeV}]}\right)^{d/2}, \label{eq:sqrtsy2}
\end{eqnarray}

%(baryon, strangeness and electric charge quantum numbers)

%
%%%%%%%%%%%%%%%%%%%%%%%%%
\begin{figure}[!htb]
\includegraphics[width=12.cm]{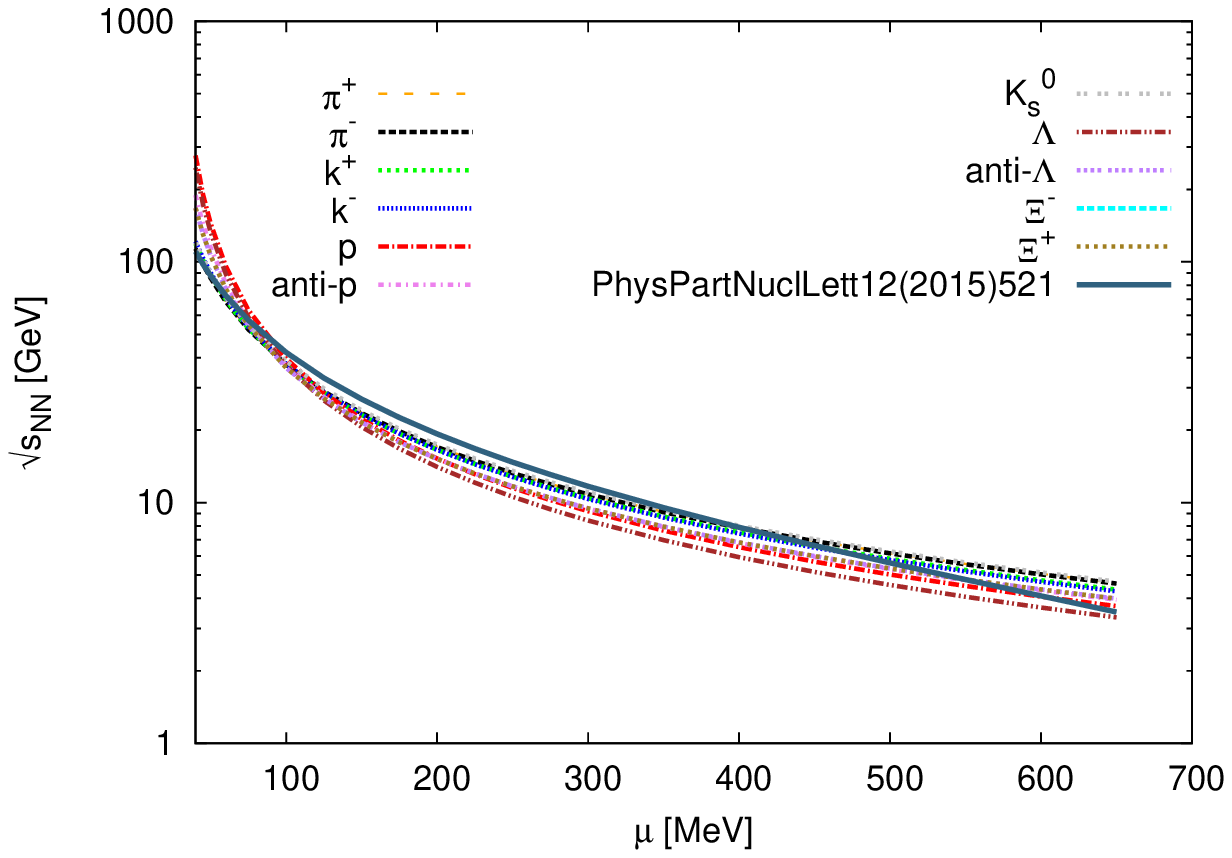}
\caption{In a semi-log scale, $\sqrt{s_{\mathtt{NN}}}~\mathtt{[GeV]}$ is presented in dependence on $\mu~\mathtt{[MeV]}$ for $\pi^+$, $\pi^-$, $K^+$, $K^-$, $p$, $\bar{p}$, $K_s^0$, $\Lambda$, $\bar{\Lambda}$, $\Xi^-$, $\Xi^+$ compared with the thermal model estimations \cite{Tawfik:2013bza,Andronic:2009qf}.  \label{fig:3}}
\end{figure}
%%%%%%%%%%%%%%%%%%%%%%%%%%%

Figure \ref{fig:3} presents the results obtained from Eq. (\ref{eq:muy}) for the eleven particles compared with the results obtained from the statistical thermal models \cite{Tawfik:2013bza} (solid curve). We observe that all particles agree well with the thermal model calculations. Such an agreement looks better than the one reported in ref. \cite{Tawfik:2019gpc}. The improvement comes from the more collision energies considered and the inclusion of more particles with more electric charge and strangeness contents. 

In ref. \cite{Tawfik:2019gpc}, the results deduced for an ensemble of $\pi^+$, $\pi^-$, $K^+$, $K^-$ were shown to have an almost identical energy dependence similar to that of the thermal models \cite{Tawfik:2013bza,Andronic:2009qf}, while for $p$ and $\bar{p}$ the energy dependence agrees well, at low energy. At high energies, both have a slightly different energy dependence. In the present calculations, all strange and charged particles seem to agree excellently with the thermal models. We conlcude that our new expressions greatly improve the results obtained in ref. \cite{Tawfik:2019gpc}. The least square fits compared with the thermal models for the present results and ref. \cite{Tawfik:2019gpc} are listed in Tab. \ref{Tab:13}.

%%%%%%%%%%%%%
\begin{table}[ht]
\centering % used for centering table
\begin{tabular}{|c|c|c|} \hline\hline
  Particle & calculated $\chi^2$ & $\chi^2$ from ref. \cite{Tawfik:2019gpc} \\
   \hline
 $\pi^+$ & $1.073$ & $1.511$ \\ \hline 
 $\pi^-$ & $0.621$ & $1.496$ \\ \hline 
 $K^-$ & $0.621$ & $1.849$ \\ \hline 
 $K^+$ & $0.787$ & $1.601$ \\ \hline 
 $p$ & $1.065$ & $0.869$\\ \hline 
 $\bar{p}$ & $0.626$ & $2.282$ \\ \hline 
 $K_s^0$ & $0.838$ & -\\  \hline 
 $\Lambda$ & $1.295$ & -\\ \hline 
 $\bar{\Lambda}$ & $1.021$ & -\\ \hline 
 $\Xi^-$ & $1.358$ & -\\ \hline 
 $\Xi^+$ & $1.509$ & -\\
 \hline
 \end{tabular}
\caption{The least square fits from the present study compared with ref. \cite{Tawfik:2019gpc}. \label{Tab:13} }
\end{table}
%%%%%%%%%%%%%

The chemical potential parameter $\mu$ in thermal model exclusively represents the baryon chemical potential $\mu_B$ only but here we mean by $\mu$ the entire chemical potential components; $\mu=n_B \mu_B + n_S \mu_S + n_Q \mu_Q$, where $n_B$, $n_S$, and $n_Q$ are baryon, strangeness, and charge quantum numbers for each particle \cite{Tawfik:2014dha,Tawfik:2017oyn}. The relevant chemical potentials, $\mu_B$, $\mu_S$, and $\mu_Q$ refer to the baryon, strangeness, and electric charge chemical potential, respectively. From the well-known quark constituents, the particles considered in this study can be classified as
\begin{itemize}
\item For $p$ and $\bar{p}$, $\mu=n_B \mu_B + n_Q \mu_Q$,
\item for $K^+$ and $K^-$, $\mu= n_S \mu_S + n_Q \mu_Q$,
\item for $\pi^+$ and $\pi^-$, $\mu=n_Q \mu_Q$,
\item for $K_s^0$, $\mu=n_S \mu_S$,
\item for $\Lambda$ and $\bar{\Lambda}$, $\mu=n_B \mu_B + n_S \mu_S$,
\item for $\Xi^+$ and $\Xi^-$, $\mu=n_B \mu_B + n_S \mu_S + n_Q \mu_Q$.
\end{itemize}

Another novel result of the present calculations, which greatly distinguishes them from the one reported in ref. \cite{Tawfik:2019gpc}, is a distinctive estimation for the various types of chemical potentials
\begin{itemize}
\item $\mu_Q = (6.073\pm1.385) + (10.694\pm0.323)\, y^2$,

\item $\mu_S = (1.276\pm3.061) + (2.247\pm0.713)\, y^2$,

\item $\mu_B = (21.571\pm3.805) + (6.674\pm0.887)\, y^2$.
\end{itemize}
As discussed, any attempt to propose separate estimations for the various types of chemical potentials is constrained by various laws conservation. Here, we have an almost-entirely empirical estimation, i.e. the experimental results are the only inputs needed to estimate any of the various chemical potentials.

We have obtained these estimations as follows. First, we substitute the well-known quantum numbers $n_Q$, $n_S$, $n_B$ for $\pi^+$, $K^+$, $p$, respectively. Then, from $\pi^+$, we find that the generic chemical potential is given by one type, the electric charge, i.e. $\mu = \mu_Q$. For $K^+$ and by using the expression just obtained for $\mu_Q$, we find that $\mu$ could be be related to $\mu_S$, only. For $\mu_B$, we have substituted the expression just obtained for $\mu_Q$ into $\mu$ for $p$. When substituting the three types of chemical potential in the remaining particles, at a given $y$, we get: 
\begin{itemize}
\item For $\Lambda$: $\mu\equiv \mu_B - \mu_S = (20.295\pm6.86) + (4.427\pm1.59)\, y^2$. 
\item For $\Xi^-$: $\mu\equiv \mu_B - 2 \mu_S - \mu_Q = (12.946\pm8.241) + (8.515\pm1.923)\, y^2$.
\end{itemize}
Approximately, the resulting generic $\mu$ agrees well with the combination of the various types of chemical potentials. 

\section{Conclusions}
\label{sec:cncl}

Exclusively based on the experimental results on $p_{\bot}$ and $d^2 N/(2 \pi p_{\bot} dp_{\bot} dy)$ for the well-identified produced particles, $\pi^+$, $\pi^-$, $K^+$, $K^-$, $p$, $\bar{p}$, we present an almost-entirely empirical estimation of the corresponding chemical potentials as functions of rapidity \cite{Tawfik:2019gpc}. In doing this, we have refined various components introduced in ref. \cite{Tawfik:2019gpc}. In the present calculations, we take into account more strangeness  contents, where additional $K_s^0$, $\Lambda$, $\bar{\Lambda}$, $\Xi^-$, $\Xi^+$ are included in. These particles bring more electric charges and baryon quantum numbers to the ensemble. Also, we cover more collision energies. This allows us to present for the first time an estimation for the various types of the chemical potentials, namely baryon, strangeness and electric charge chemical potentials as functions of the rapidity. 

The main result obtained is a universal approach relating the chemical potential with the rapidity for all produced particles; $\mu=a+b y^2$, where $a$ and $b$ are constants. An excellent agreement was also found, when comparing the energy dependence of the chemical potential; $\sqrt{s_{\mathtt{NN}}}=c[(\mu-a)/b]^{d/2}$, where $c$ and $d$ are constants to be fixed from phenomenological observations such as the statistical thermal models. 

We found that the results obtained agree well with the energy dependence of $\mu$ based on the statistical thermal approach for an ideal gas of hadron resonances \cite{Tawfik:2013bza,Andronic:2005yp}. In ref. \cite{Tawfik:2019gpc}, it was concluded that the results deduced for an ensemble of $\pi^+$, $\pi^-$, $K^+$, $K^-$ have an almost identical energy dependence similar to the one of the statistical thermal models \cite{Tawfik:2013bza,Andronic:2009qf}, while for $p$ and $\bar{p}$ it was found that the energy dependence matches only, at low energies. At high energies, both particles seem to have a slightly different energy dependence. In the present calculations, we found that all strange and charged particle agree excellently with the statistical thermal models. Such an improvement relative ref. \cite{Tawfik:2019gpc} is manifold. 

Besides an excellent agreement with the statistical thermal models, the present approach distinguishes between the various types of chemical potential; $\mu=n_B \mu_B + n_S \mu_S + n_Q \mu_Q$, where $n_B$, $n_S$, and $n_Q$ are baryon, strangeness, and charge quantum numbers for each of the particles \cite{Tawfik:2014dha,Tawfik:2017oyn}. The present approach presents separate estimations for $\mu_B$, $\mu_S$, and $\mu_Q$, the baryon, the strangeness, and the charge chemical potential, respectively. 

\bibliographystyle{aip}

\bibliography{strangeMuY}

\begin{thebibliography}{10}

\bibitem{Tawfik:2000mw}
A.~M. Tawfik and E.~Ganssauge,
\newblock Acta Phys. Hung. {\bf A12}, 53 (2000).

\bibitem{Heinz:2000ba}
U.~W. Heinz,
\newblock Nucl. Phys. {\bf A685}, 414 (2001).

\bibitem{Gyulassy:2004zy}
M.~Gyulassy and L.~McLerran,
\newblock Nucl. Phys. {\bf A750}, 30 (2005).

\bibitem{Heinz:2011kt}
U.~Heinz, C.~Shen, and H.~Song,
\newblock AIP Conf. Proc. {\bf 1441}, 766 (2012).

\bibitem{Adamczyk:2013dal}
L.~Adamczyk and others (STAR~Collaburation),
\newblock Phys. Rev. Lett. {\bf 112}, 032302 (2014).

\bibitem{Ryu:2017qzn}
S.~Ryu et~al.,
\newblock Phys. Rev. {\bf C97}, 034910 (2018).

\bibitem{Bzdak:2019pkr}
A.~Bzdak et~al.,
\newblock (2019).

\bibitem{Tawfik:2004vv}
A.~Tawfik,
\newblock J. Phys. {\bf G31}, S1105 (2005).

\bibitem{Tawfik:2004sw}
A.~Tawfik,
\newblock Phys. Rev. {\bf D71}, 054502 (2005).

\bibitem{Tawfik:2019yxn}
A.~N. Tawfik, M.~Maher, A.~H. El-Kateb, and S.~Abdelaziz,
\newblock (2019).

\bibitem{Tawfik:2018sji}
A.~N. Tawfik, H.~Yassin, and E.~R.~A. Elyazeed,
\newblock EPL {\bf 126}, 41001 (2019).

\bibitem{Tawfik:2014eba}
A.~N. Tawfik,
\newblock Int. J. Mod. Phys. {\bf A29}, 1430021 (2014).

\bibitem{Tawfik:2016jzk}
A.~Tawfik, M.~Y. El-Bakry, D.~M. Habashy, M.~T. Mohamed, and E.~Abbas,
\newblock Int. J. Mod. Phys. {\bf E25}, 1650018 (2016).

\bibitem{Tawfik:2013eua}
A.~Tawfik,
\newblock Phys. Rev. {\bf C88}, 035203 (2013).

\bibitem{Tawfik:2013dba}
A.~Tawfik,
\newblock Nucl. Phys. {\bf A922}, 225 (2014).

\bibitem{Tawfik:2012si}
A.~Tawfik,
\newblock Adv. High Energy Phys. {\bf 2013}, 574871 (2013).

\bibitem{Tawfik:2005qn}
A.~Tawfik,
\newblock Nucl. Phys. {\bf A764}, 387 (2006).

\bibitem{Tawfik:2004ss}
A.~Tawfik,
\newblock Europhys. Lett. {\bf 75}, 420 (2006).

\bibitem{Tawfik:2019gpc}
A.~N. Tawfik, M.~A. Wahab, H.~Yassin, and H.~N.~E. Din,
\newblock (2019).

\bibitem{Tawfik:2014dha}
A.~N. Tawfik, M.~Y. El-Bakry, D.~M. Habashy, M.~T. Mohamed, and E.~Abbas,
\newblock Int. J. Mod. Phys. {\bf E24}, 1550067 (2015).

\bibitem{Adamczyk:2017iwn}
L.~Adamczyk and others (STAR~Collaboration),
\newblock Phys. Rev. {\bf C96}, 044904 (2017).

\bibitem{Adam:2019koz}
J.~Adam and others (STAR~Collaburation),
\newblock (2019).

\bibitem{Aggarwal:2010ig}
M.~M. Aggarwal and others (STAR~Collaburation),
\newblock Phys. Rev. {\bf C83}, 024901 (2011).

\bibitem{Abelev:2006cs}
B.~I. Abelev and others (STAR~Collaburation),
\newblock Phys. Rev. {\bf C75}, 064901 (2007).

\bibitem{Abelev:2008ab}
B.~I. Abelev and others (STAR~Collaboration),
\newblock Phys. Rev. {\bf C79}, 034909 (2009).

\bibitem{Tawfik:2013bza}
A.~N. Tawfik and E.~Abbas,
\newblock Phys. Part. Nucl. Lett. {\bf 12}, 521 (2015).

\bibitem{Andronic:2009qf}
A.~Andronic, P.~Braun-Munzinger, and J.~Stachel,
\newblock Acta Phys. Polon. {\bf B40}, 1005 (2009).

\bibitem{Tawfik:2017oyn}
A.~N. Tawfik, H.~Yassin, and E.~R. Abo~Elyazeed,
\newblock Phys. Part. Nucl. Lett. {\bf 14}, 843 (2017).

\bibitem{Andronic:2005yp}
A.~Andronic, P.~Braun-Munzinger, and J.~Stachel,
\newblock Nucl. Phys. {\bf A772}, 167 (2006).

\end{thebibliography}

\end{document}